\documentclass[%
 aip,apl,
 amsmath,amssymb,
 reprint,%
]{revtex4-1}

\usepackage{graphicx}
\usepackage{dcolumn}
\usepackage{bm}

\usepackage[utf8]{inputenc}
\usepackage[T1]{fontenc}
\usepackage{mathptmx}
\usepackage{etoolbox}

\usepackage{braket}

\makeatletter
\def\@email#1#2{%
 \endgroup
 \patchcmd{\titleblock@produce}
  {\frontmatter@RRAPformat}
  {\frontmatter@RRAPformat{\produce@RRAP{*#1\href{mailto:#2}{#2}}}\frontmatter@RRAPformat}
  {}{}
}%
\makeatother
\begin{document}

\preprint{AIP/123-QED}

\title[Title]{Quasi-Continuous Cooling of a Microwave Mode on a Benchtop using Hyperpolarized NV$^-$ Diamond}
\author{Wern Ng}
   \affiliation{%
Department of Materials, Imperial College London, South Kensington SW7 2AZ, London, United Kingdom}%
  \email{kuok.ng14@imperial.ac.uk}
\author{Hao Wu}%

\affiliation{ 
Center for Quantum Technology Research and Key Laboratory of Advanced Optoelectronic Quantum Architecture and Measurements (MOE), School of Physics, Beijing Institute of Technology, Beijing 100081, China
}%

\affiliation{ 
Beijing Academy of Quantum Information Sciences, Beijing 100193, China
}%

\author{Mark Oxborrow}
   \affiliation{%
Department of Materials, Imperial College London, South Kensington SW7 2AZ, London, United Kingdom}%

 \email{m.oxborrow@imperial.ac.uk}
\date{\today}

\begin{abstract}

We demonstrate the cooling of a microwave mode at 2872 MHz through its interaction with optically spin-polarized NV$^-$ centers in diamond at zero applied magnetic field, removing thermal photons from the mode.
By photo-exciting (pumping) a brilliant-cut red diamond jewel with a continuous-wave 532-nm laser, outputting 2 W,
the microwave mode is cooled down to a noise temperature of 188 K.
This noise temperature can be preserved continuously for as long as the diamond is optically excited and kept cool. The latter requirement restricted operation out to 10~ms in our preliminary setup.  
The mode-cooling performance of NV$^-$ diamond is directly compared against that of pentacene-doped \textit{para}-terphenyl, where we find that the former affords the advantages of cooling immediately upon light excitation without needing to mase beforehand (or at all) and being able to cool
continuously at substantially lower optical pump power.

\end{abstract}

\maketitle

A century ago, the invention of superheterodyning substantially enhanced the sensitivity with which weak radio signals could be detected in the face of noise. Today, after many decades of research into semiconducting materials and devices, low-noise amplifiers operating at GHz frequencies with large fractional bandwidths can be readily purchased. These amplifiers, 
typically in the form of HEMT-based MMICs, require only a source of low-voltage d.c.~power to operate and offer noise figures as low as  $\sim$0.3~dB\cite{Weinreb1982}, corresponding to amplifier noise temperatures of just a few tens of kelvin. This fantastic technological achievement has left us in the situation where the sensitivity at which rf measurements can be made at room temperature is almost entirely limited by thermal (i.e.~Johnson-Nyquist) noise. The obvious route to further improving sensitivity and attaining the single-photon (a.k.a. ``quantum'') limit is to cool the entire instrument/experiment down within a refrigerator. For measurements at microwave frequencies, this means cooling to mK temperatures by means of a dilution refrigerator. However, the machinery is bulky, fragile, and power hungry (consuming kilowatts, overall) whilst offering minuscule cooling powers (milliwatts, at best). Alas, nobody has yet managed to miniaturize a dilution refrigerator to fit and operate (off batteries) within the form factor of a mobile phone or wrist watch. 

A recent paper\cite{HaoWu2021} explored a radical alternative to physical refrigeration of the instrument. Instead, a single microwave mode is cooled, reducing the number of thermal microwave photons occupying it and thus reducing the amount of Johnson noise, in units of watts per Hz of bandwidth, extracted from it. This more targeted form of cooling was achieved through the mode's interaction with a strongly spin-polarized material exhibiting a cryogenic spin temperature across a paramagnetic transition in tune with the mode. The material in question was an organic molecular crystal, namely pentacene-doped \textit{p}-terphenyl (Pc:PTP), whose spin-coldness was generated by optical pumping. This demonstrator suffered from initially masing after the application of a high-power (and high overall energy) optical pulse before its mode-cooling effects could be accessed, as well as only being able to operate in pulsed mode, cooling the microwave mode for just a few hundreds of microseconds at a time. In this Letter, we report on the sustained cooling of a microwave mode, achieved using photoexcited negatively charged nitrogen vacancies (NV$^-$) in diamond as the cryo-spin-polarized absorber instead. The setup is run entirely at zero d.c. applied magnetic field (ZF) and under ambient temperature on the lab benchtop. Because the setup operates at ZF, there is no need to set up strong magnets or align the diamond crystal to a magnetic field (no orientation dependence). NV$^-$ diamond differs from Pc:PTP in many ways; most pertinently, it requires substantially lower optical pumping power to become spin-polarized to an equivalent extent. In this study, we describe the mechanism, anatomy and performance of our new mode cooler and compare the relative merits of NV$^-$ diamond to Pc:PTP in this application.

Our experiment uses a round brilliant cut diamond with a girdle diameter of 2.65 mm, a table size of 2 mm, a total height of 1.5 mm, and a pavilion angle of 41.5$^\circ$. Though natural (i.e., mined from the ground), the diamond had been ``color enhanced'' through annealing. It has a dark red color and fluoresces bright red under 532 nm illumination. From UV/Vis spectroscopy shown in the Supplementary Material (SM), the sample absorbed broadly at 576 nm and sharply at 638 nm, the latter being the characteristic zero-phonon line of NV$^-$ centers. Using the absorption coefficient and an estimate of the absorption cross-section near 638 nm,\cite{Fraczek17} the concentration of NV$^-$ in the diamond is estimated to be  $6\times10^{17}$ cm$^{-3}$.
\begin{figure*}[ht!]
\includegraphics{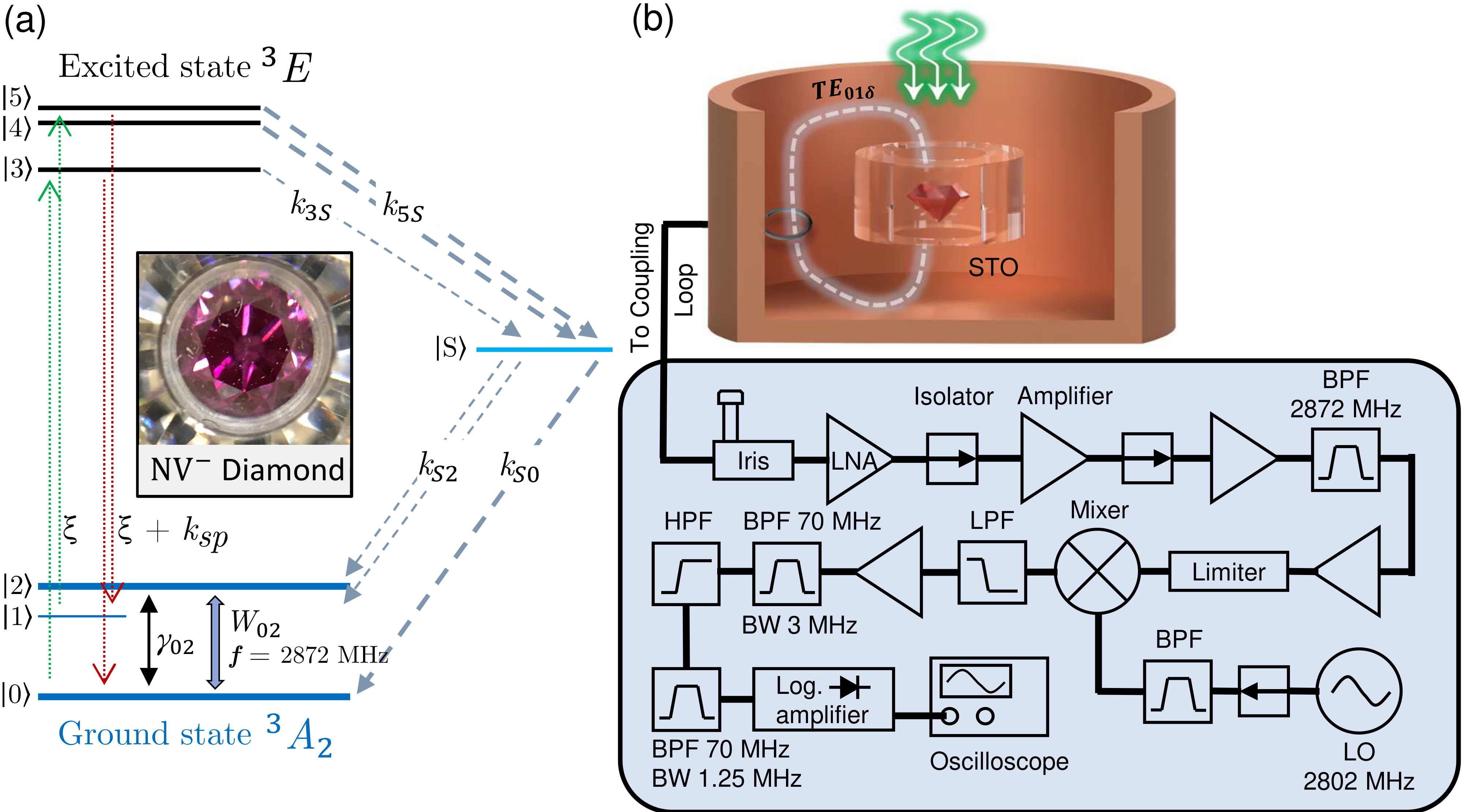}
\caption{\label{fig:jablon}(a) Jablonski diagram showing photoexcitation of NV$^-$ centers using 532 nm light and subsequent relaxations to the metastable singlet state and $^3A_2$ state through ISC, through the rate constants $k_{3S},k_{5S},k_{S2},k_{S0}$. The spin-lattice relaxation rate $\gamma_{02}$, stimulated transition rate $W_{02}$ and optical pumping parameter $\xi$ are also indicated. (b) Cross-sectional render of the copper cavity housing the NV$^-$ diamond within a STO dielectric resonator, with TE$_{01\delta}$ mode tuned to $f_{\text{mode}}=2872$ MHz and 532 nm excitation from above. The cooling response is measured through a coupling loop connected to the heterodyne receiver setup (schematic illustrated below). LPF, HPF and BPF stand for low-, high- and band-pass filters respectively.}
\end{figure*}

\begin{figure}
\includegraphics[width=0.4\textwidth]{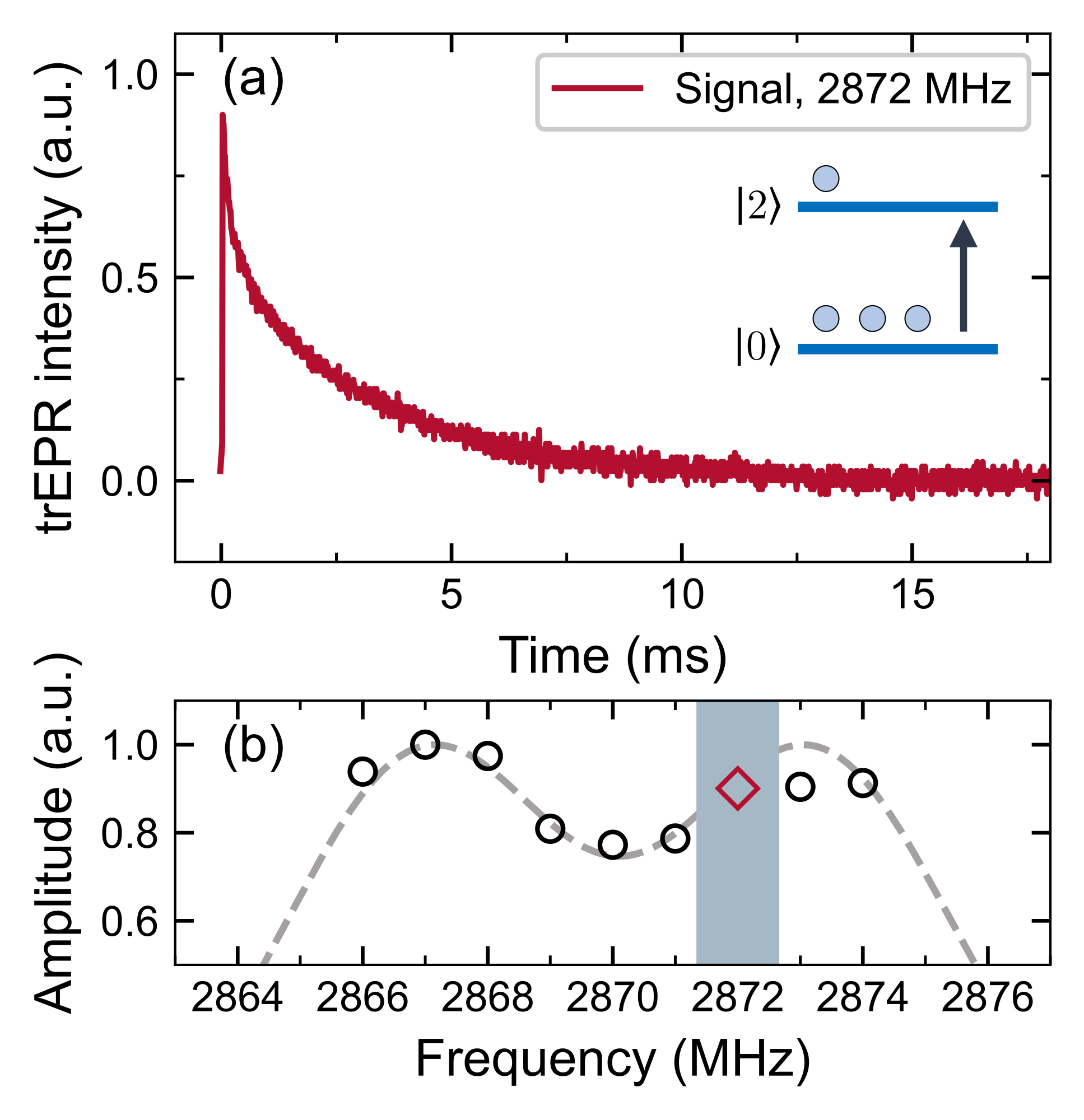}
\caption{\label{fig:trepr} (a) ZF-trEPR of NV$^-$ diamond, with the $\ket{0}\rightarrow\ket{2}$ transition illustrated. The population (number of circles) is higher in $\ket{0}$, indicating an absorptive signal. (b) Peak amplitudes of the ZF-trEPR responses at different microwave frequencies. $f_{\text{mode}}=2872$ MHz is indicated by a red marker, and the shaded region shows the SAW filter bandwidth.}
\end{figure}
Upon illumination with 532-nm laser light, the NV$^-$ centers will be excited from the triplet ground state ($^3A_2$) to the first triplet excited state ($^3E$) . They then undergo intersystem crossing (ISC) into a metastable singlet state via spin-dependent rates ($k_{3S},k_{5S}$) and again undergo ISC back to $^3A_2$ with rates $k_{S2},k_{S0}$ (see Fig.~\ref{fig:jablon}(a))\cite{Tetienne2012}. The optical pumping rate is expressed by the pumping parameter $\xi$, which is proportional to the instantaneous laser power coupled into the sample. A full explanation of the dynamics is given in the SM. 

At zero applied d.c.~magnetic field, the frequency corresponding to the energy difference between the lowest energy level ($\ket{0}$, corresponding to $m_s=0$) and the upper pair of degenerate levels ($\ket{1},\ket{2}$, corresponding to $m_s=\pm1$) is 2870~MHz\cite{Tetienne2012}.  The spin-dependency of the ISC rates acts to increase the number of NV$^-$ centers in the $\ket{0}$ level relative to the $\ket{1}$ and $\ket{2}$ levels\cite{dia_penta_review}, so causing the transitions between 
$\ket{1}$ or $\ket{2}$ and $\ket{0}$ to become absorptively hyperpolarized\cite{Tetienne2012,dia_penta_review}. Additionally, the central microwave resonance at 2870 MHz for a bulk sample of NV$^-$ centers in diamond will be split into two distinct resonance peaks due to the effects of local electric fields\cite{Mittiga2018}. Measurements of our sample using a homemade photoexcited transient electron paramagnetic resonance spectrometer at zero field\cite{HaoWu2019} (ZF-trEPR) confirm  this splitting of the resonance line with the strongest observed ZF-trEPR signals in the vicinities of 2867 MHz and 2873 MHz; see Fig.~\ref{fig:trepr}. Similar resonance splittings have been observed in zero-field ODMR of NV$^-$ diamond\cite{Acosta2010,Mittiga2018}. 

\begin{figure*}
\includegraphics{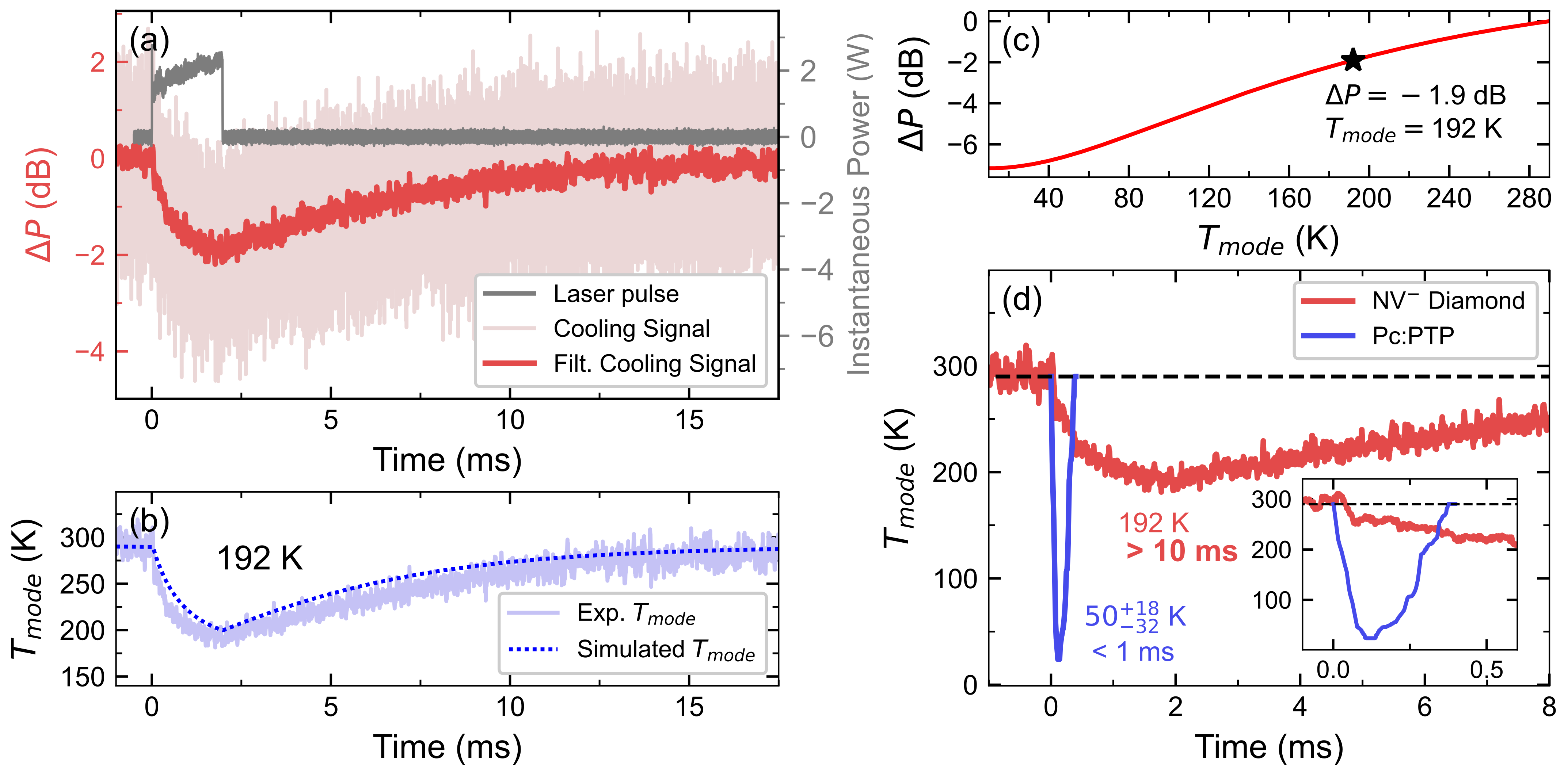}
\caption{\label{fig:dia_sig}(a) Cooling response of NV$^-$ diamond measured at $f_{\text{mode}}=2872$ MHz. The light red plot is the signal average of 63 individual scans, and the dark red plot is the signal average after digital filtering that more clearly shows the depth of the power reduction. The grey plot shows the photodiode response of the square laser pulse used to excite the diamond, converted to power in watts. (b) Calculated experimental $T_{\text{mode}}$ achieved from the observed filtered power reduction signal. The simulated $T_{\text{mode}}$ from solving the optical and spin dynamics equations of NV$^-$ diamond is plotted as the dotted line. The cooling signal reaches a depth of 192 K. (c) Simulation showing the relationship between the power reduction $\Delta P$ and cooling of $T_{\text{mode}}$, with the point corresponding to the minimum $\Delta P$ from (a) marked. (d) Comparison of the cooling response between NV$^-$ diamond and Pc:PTP, where the response of Pc:PTP is adapted from Ref.~\onlinecite{HaoWu2021}. The inset plot shows the same signals but at a shorter time scale.}
\end{figure*}

Mode cooling can be performed by tuning a sufficiently high-Q microwave mode to the frequency of either of the two absorptive peaks. We measured the mode-cooling response at the same discrete nine interogating frequencies as used for ZF-trEPR and found that the cooling was deepest when the microwave mode was tuned to 2872 MHz. The mode cooling at 2867 MHz was not quite as deep, but far better than that at 2870 MHz. We thereupon chose $f_{\text{mode}}=2872$ MHz as the frequency at which to study the cooling in detail; this is indicated as the transition frequency of $\ket{0}\rightarrow\ket{2}$ in Fig.~\ref{fig:jablon}(a).

In a similar manner to how the $X\rightarrow Z$ triplet transition of Pc:PTP in our previous work was used to cool a microwave mode at 1450 MHz\cite{HaoWu2021}, we here exploit the $\ket{0}\rightarrow\ket{2}$ transition in NV$^-$ diamond for mode cooling. We use a similar setup as in the previous work, incorporating a high-gain superheterodyne receiver to measure the instantaneous microwave power extracted from the microwave mode in a cavity, monitored on an oscilloscope (LeCroy DDA-260, DC-coupled) triggered by the optical pumping pulse \cite{HaoWu2021}. A few modifications are worth noting: to be resonant at the higher frequency of 2872 MHz, the STO dielectric resonator is now almost half of its previous size (one eighth in volume), with a lower loaded quality factor ($Q_L$) of 2900. A microwave iris (stub-tuner) has been added between the resonator and LNA to allow precise matching of the cavity to critical coupling (reflection coefficient $\Gamma^0_c=0$). The optical pump source is a 2-W continuous-wave 532-nm diode-pumped Nd:YAG solid-state laser, whose output is gated using square pulses from a pulse generator. The NV$^-$ diamond, placed inside the STO resonator within a copper cavity as depicted in Fig.~\ref{fig:jablon}(b), is excited by the laser from above through a hole in the copper cavity, with the beam focused down to a spot diameter of 1.5 mm at the sample. Finally, since the lower loaded $Q_L$ corresponds to a wider -3 dB linewidth of 1 MHz, the original 50-kHz-bandwidth SAW bandpass filter used before the detector in our previous work\cite{HaoWu2021} is replaced with a 1.25-MHz-bandwidth version; this avoids excessive temporal delay and distortion of the detected power signal (albeit at the expense of allowing a small fraction of noise on the shoulder's of the resonator's line profile into the superheterodyne signal path). As before, no attempt is made to reject image noise through more sophisticated schemes (such as double-conversion superhet). As seen in Fig.~\ref{fig:trepr}(b), the resonator's linewidth is narrow enough to ensure that only the NV$^-$ centers resonant at 2872 MHz, to which the resonator is tuned, get excited (and not those at 2867 MHz).  The complete receiver arrangement is shown in Fig.~\ref{fig:jablon}(b).

Fig.~\ref{fig:dia_sig}(a) shows the resultant reduction in noise power at the superheterodyne receiver's output when the NV$^-$ diamond is excited with a 2-ms square pulse of 532-nm laser light with a repetition frequency of 6 Hz and instantaneous power of 2 W. The signal is averaged over 63 separate (statistically independent) measurements, after which it is smoothed using a median filter to compensate for the higher SAW filter bandwidth and to more clearly show the depth of the cooling response. This measurement indicates a reduction in the noise power of $-1.9$ dB, corresponding to a drop
of $\sim$100~K in the microwave mode's temperature from $T_{\text{mode}}=290$ K to  $T_{\text{mode}}=192$ K; see Fig.~\ref{fig:dia_sig}(c). The equivalent drop in the number of thermal photons $q$ occupying the mode can be calculated through Eq.~\ref{eq:photon}.
\begin{eqnarray}
q=\left(\exp(hf_{\text{mode}}/k_BT_{\text{mode}})-1\right)^{-1}
\label{eq:photon}.
\end{eqnarray}
In Eq.~\ref{eq:photon}, $f_{\text{mode}}$ is the mode frequency at 2872 MHz. Cooling the mode down to 192 K represents a reduction from 2103 to 1392 in the number of thermal photons.

The cooling signal then decays back to the baseline noise power (indicated by $\Delta P=0$) with a decay time of about 10 ms after the laser pulse ends. The temperature decrease and time dynamics of the signal were successfully simulated in Fig.~\ref{fig:dia_sig}(b), where $T_{\text{mode}}$ is simulated from solving coupled differential equations dictating NV$^-$ diamond's spin dynamics and $q$ (see SM)\cite{Tetienne2012,Hintze2017,Barry2020,Salvadori2017,siegman1964microwave,qorvoamp,kraus1986,Tse-Luen2007}.

We here compare the mode-cooling capabilities of NV$^-$ diamond to those of Pc:PTP as directly as possible. 
The former has an obvious advantage in that it immediately cools the microwave mode without first masing at zero field \cite{HaoWu2021}. In contrast to Pc:PTP, NV$^-$ diamond exhibits no cross-over (from absorptive to emissive) at later times; it can only cool at zero field. In Fig.~\ref{fig:dia_sig}(d), the cooling signal  from our present experiment is compared against an adaption of a measured cooling signal for Pc:PTP (post-masing) reported in our previous work \cite{HaoWu2021} where Pc:PTP is shown to be able to cool the mode to a much lower temperature (50 K), but requires high-power pulsed excitation lasting only 300 $\mu$s at $\sim$5-kW peak pulse power. Due to the pulsed nature of the excitation, the cooling response of Pc:PTP does not last more than 0.5 ms. Conversely, the cooling afforded by NV$^-$ diamond, even from a pulsed excitation, lasts up to 10 ms which is longer than that attained with Pc:PTP by two orders of magnitude. Furthermore, compared to the dye laser used for pumping Pc:PTP, the 2-W Nd:YAG laser used here is substantially cheaper and easier to purchase and maintain. The disadvantage of NV$^-$ diamond though is its inferior depth of the cooling. Methods to overcome this will be discussed towards the end of this Letter.

Attempts at pumping Pc:PTP with the same 2-ms 532-nm laser pulse used for NV$^-$ diamond (after switching back to an STO resonator tuned to 1450 MHz) produced no cooling response. Likewise, we attempted to excite NV$^-$ diamond using Q-switched 5.5 ns high power pulses from an optical parametric oscillator (Litron Aurora II Integra) at 532 nm. This gave a much smaller power reduction of only 0.4 dB, which corresponds to much weaker cooling with $T_{\text{mode}}$ dropping by 30 K only. We note that NV$^-$ diamond produces a lower polarization if pumped with pulse lengths shorter than the decay times of $\ket{S}$ to $^3A_2$ ($\sim1\ \mu$s)\cite{dia_penta_review}, so this may explain why the nanosecond pump pulses produce a poor cooling result. Overall, the mode cooling afforded by Pc:PTP is deep but short, whereas that afforded by NV$^-$ diamond is not as deep but long, while requiring vastly lower instantaneous optical pump power.

\begin{figure}
\includegraphics{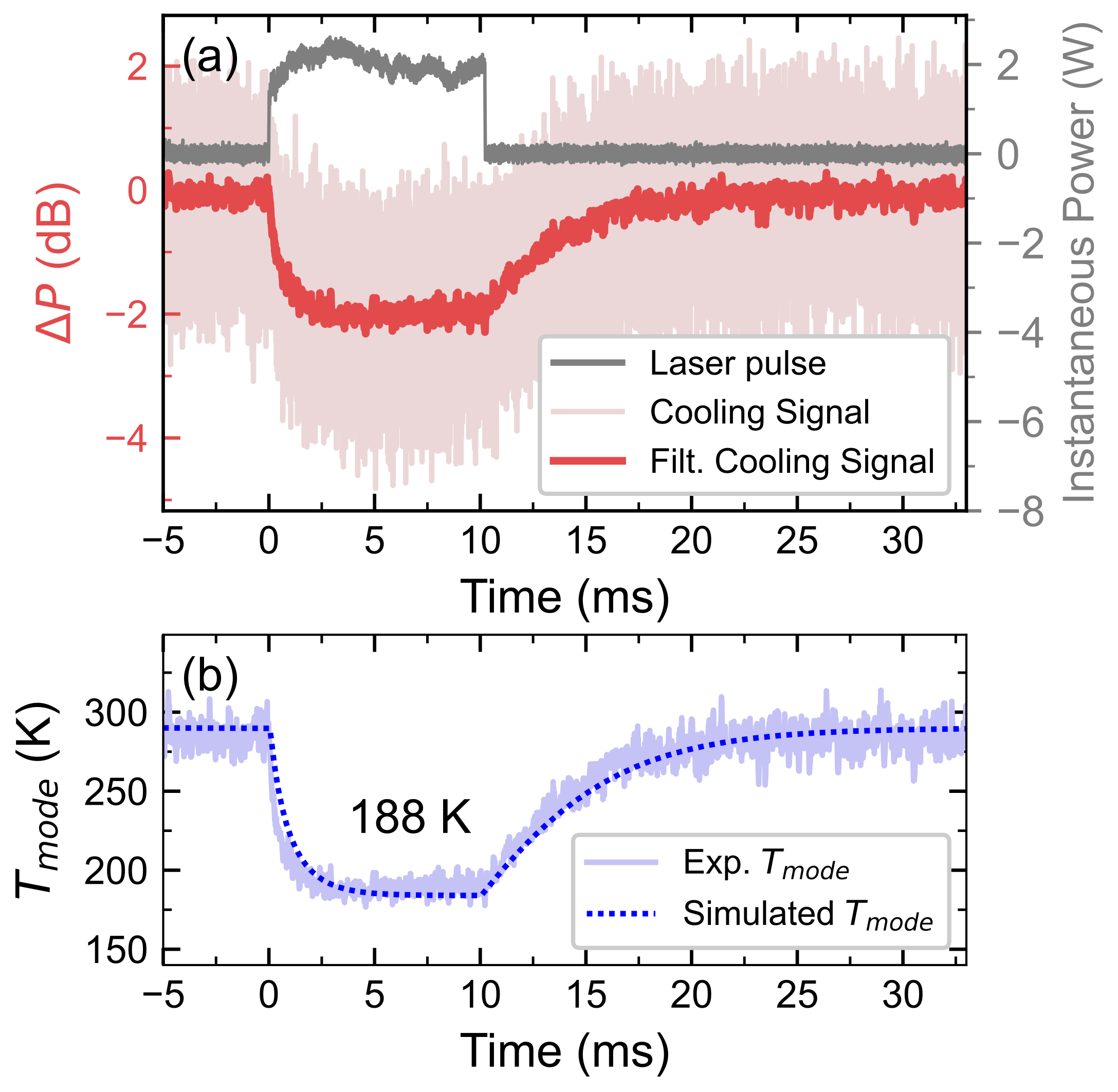}
\caption{\label{fig:cont} (a) Cooling response of NV$^-$ diamond measured at $f_{\text{mode}}=2872$ MHz when excited with a long 10 ms laser pulse. The light red plot is again the signal average of 63 individual scans, and the dark red plot is that signal after digital filtering. The power reduction is seen to hold steadily for as long as the laser excitation is present. (b) Calculated experimental $T_{\text{mode}}$ achieved from the observed filtered power reduction signal. The continuous cooling of the microwave mode was also successfully simulated as the dotted line. The cooling signal reaches a lower depth of 188 K and maintains it for the 10 ms of laser excitation.}
\end{figure}

After having demonstrated how NV$^-$ diamond could cool the microwave mode under pulsed excitation, we investigated whether the microwave mode could be cooled continuously for as long as the NV$^-$ diamond was optically pumped, beyond its relaxation time (about 10 ms in Fig.~\ref{fig:dia_sig}(a)). Fig.~\ref{fig:cont}(a) shows the noise power reduction when the NV$^-$ diamond is pumped by a 10-ms square laser pulse (repetition frequency of 6 Hz). The resultant signal can be seen to follow the time profile of the laser excitation, where the mode is maintained at its cooled temperature for as long as the laser pulse is on. From Fig.~\ref{fig:cont}(b), the experimental $T_{\text{mode}}$ mode is calculated to be cooled down to and held at the temperature of 188 K (reducing the number of thermal photons to 1363). After the laser pulse, the system relaxes with the same decay time as in Fig.~\ref{fig:dia_sig}(a). We were similarly able to simulate the dynamics of $T_{\text{mode}}$ in Fig.~\ref{fig:cont}(b) with identical parameters to those used for simulating Fig.~\ref{fig:dia_sig}(b), but with a longer laser pulse. This then demonstrates the capability of the  material to continuously cool a microwave mode at zero field so long as it is spin-polarized (through optical excitation). Being able to continuously cool a microwave mode is a significant advantage that NV$^-$ diamond has over Pc:PTP, which can only cool in a pulsed manner.

Though we have demonstrated how NV$^-$ diamond can be used to continuously cool a microwave mode, improvements in performance are needed for the effect to be more usefully exploited.  Using a bigger diamond cut in the form of a cylindrical rod (4-mm in diameter) could improve upon the resonator's magnetic filling factor. Combined with increasing the concentration of NV$^-$ centers and boosting the excitation intensity of the pump laser, we judge that the number of spin-polarized NV$^-$ centers interacting with the TE$_{01\delta}$ could be straightforwardly increased by at least an order of magnitude. A rod shaped sample (with polished flat end windows) would prevent excessive retro-reflection of pumping light compared to the brilliant cut of our current sample, which is expressly designed to reflect light for visual appeal. Trade-offs and diminishing returns admittedly lurk here:
(i) higher concentrations of NV$^-$ necessitate higher concentrations of substitutional nitrogen defects\cite{Barry2020}, which have dipolar coupling interactions with NV$^-$ centers leading to line broadening and shorter $T_2^*$\cite{Barry2020,Kleinsasser2016,Wyk_1997}; (ii) higher optical pump power would increase polarization, yet carry the risk of converting NV$^-$ centers into NV$^0$ centers\cite{Manson2005}, so reducing the number of the former available for mode cooling.
We note that, empirically, the insertion of an ND filter across the laser's pump beam (so as to attenuate it) would always reduce the observed depth of mode cooling, suggesting that our current setup would have immediately benefited from greater pump power (up to a point).

The cooling of the microwave mode could be made truly continuous by actively cooling the diamond through one or a combination of: (i) solid thermal anchorage (heat-sinking), (ii) forced air, (iii) immersive liquid cooling (locating the diamond within a`flow tube'), or (iv) heat pipes, which are all still less costly to implement and maintain than dilution refrigeration. Given the high thermal diffusivity of diamond and the relatively modest absorbed optical pump power in need of being removed, the engineering challenge of constructing a suitable cooling system appears highly feasible. 

In conclusion, we have demonstrated how NV$^-$ diamond at zero field can be used to cool a microwave mode down to 188 K and hold it at that temperature continuously for as long as the diamond is optically excited and kept cool. The mode cooling performance of NV$^-$ diamond was then compared to that of Pc:PTP, where NV$^-$ diamond demonstrates the advantages of cooling continuously, requiring much lower excitation power, and cooling the mode immediately upon excitation to boot (in contrast to Pc:PTP which mases before cooling). Though in our current implementation, NV$^-$ diamond could not cool the mode to as low a temperature as Pc:PTP, we propose multiple avenues by which the cooling performance of NV$^-$ diamond could be improved.

\begin{acknowledgments}
We thank Ben Gaskell of Gaskell Quartz Ltd (London) for making the strontium titanate ring used. This work was supported by the U.K. Engineering and Physical Sciences Research Council through grants No. EP/K037390/1 and No. EP/M020398/1. H.W. acknowledges financial support from the China Postdoctoral Science Foundation under Grant No. YJ20210035.
\end{acknowledgments}

\section*{Data Availability Statement}

The data that support the findings of this study are available within the article [and its supplementary material].

\end{document}


\preprint{}

\title[Title]{Supplementary Material: Quasi-Continuous Cooling of a Microwave Mode on a Benchtop using Hyperpolarized NV$^-$ Diamond}
\author{Wern Ng}
   \affiliation{%
Department of Materials, Imperial College London, South Kensington SW7 2AZ, London, United Kingdom}%
  \email{kuok.ng14@imperial.ac.uk}
\author{Hao Wu}%

\affiliation{ 
Center for Quantum Technology Research and Key Laboratory of Advanced Optoelectronic Quantum Architecture and Measurements (MOE), School of Physics, Beijing Institute of Technology, Beijing 100081, China
}%
\affiliation{ 
Beijing Academy of Quantum Information Sciences, Beijing 100193, China
}%
\author{Mark Oxborrow}
   \affiliation{%
Department of Materials, Imperial College London, South Kensington SW7 2AZ, London, United Kingdom}%

 \email{m.oxborrow@imperial.ac.uk}
\date{\today}

\maketitle

\section{Additional Experimental Details}
\begin{figure*}[h!]
\includegraphics[width=0.6\textwidth]{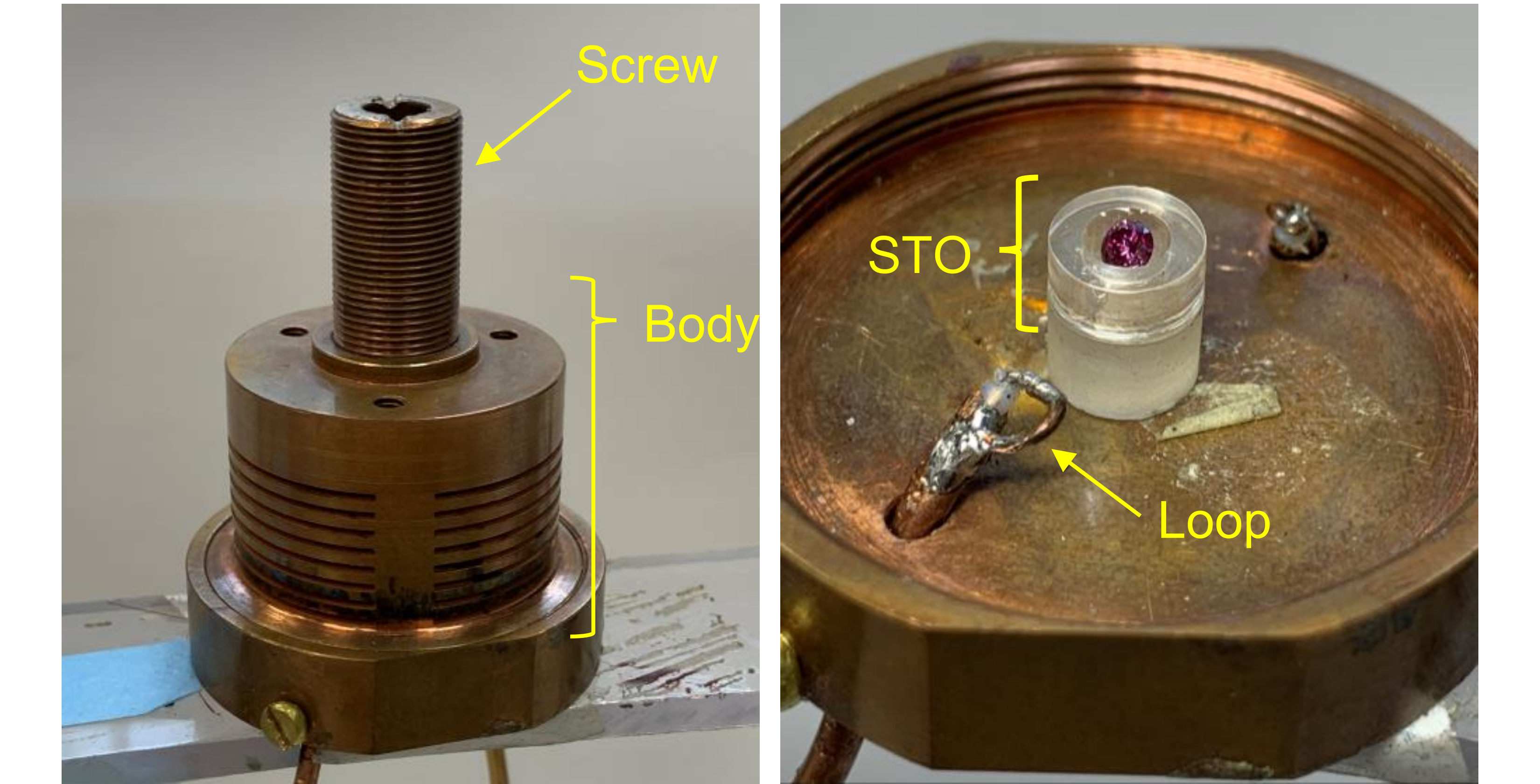}
\caption{\label{fig:res}Left image shows the copper cavity with tuning screw and `body'. The latter could be unscrewed to open the cavity. The right image shows the `body' taken off to reveal the STO resonator (on top of the polystyrene stand), diamond and coupling loop inside.}
\end{figure*}

The microwave cavity consisted of a cylindrical copper cavity (30 mm ID, 20.4 mm maximum internal height) housing a cylindrical dielectric ring of monocrystalline strontium titanate (7.27 mm OD, 4 mm ID, 3.5 mm height, all surfaces finely polished). The ring was placed on a post made of cross-linked polystyrene (similar to Rexolite) that propped the ring to a height of 3 mm above the floor of the copper cavity. The operating frequency $f_{\text{mode}}$ of the TE$_{01\delta}$ mode of the microwave cavity could be tuned by adjusting the internal height of the copper cavity. This was achieved via a `tuning' screw at the top of the copper cavity which lowered/raised a copper disk acting as a ceiling within the cavity (see Fig.~S~\ref{fig:res}). The ceiling could be adjusted so that the internal height would be anywhere between 6.5 mm and 20.4 mm (the former being the combined height of the polystyrene post and dielectric ring). The tuning screw was hollow so as to allow light to enter the cavity from above and excite the diamond sample placed in the middle of the STO ring. The loaded $Q$ ($Q_L$) of the cavity was measured to be 2900 through S$_{21}$ transmission of the cavity, measured using a VNA (HP-8753A). Zero-field transient electron paramagnetic resoance (ZF-trEPR) was carried out using a setup similar to previous literature\cite{HaoWu2019}, but with a few differences. The resonator used was an opaque dielectric ceramic ring resonator (quality factor of 900, the material having been extracted from an Adafruit GPS antenna) housed within the same copper cavity described above, instead of the STO ring. The incident microwave power for ZF-trEPR was -25 dBm, and the optical excitation used was an OPO (Litron Aurora II Integra) set to an output wavelength of 532 nm, repetition rate of 10 Hz and pulse length of 5.5 ns.

The heterodyne receiver setup used is similar to that in our previous work\cite{HaoWu2021}. It uses a cascade of amplifiers (total gain $\sim100$ dB) to achieve sufficient sensitivity. The output is fed through a narrow-bandpass SAW filter (centered at 70 MHz, with steep-skirted 1.25 MHz bandwidth) into a logarithmic detector (Pascall, SDB-7040) possessing 80 dB dynamic range. All components were used in ambient lab conditions. Digital filtering of the NV$^-$ diamond signals in all figures was done using a median filter (Python function \textit{scipy.signal.medfilt}, window size of 81) on the averaged signal data. The Pc:PTP signal in Fig.~3(d) in the main text was adapted from a single scan that was part of the cooling signal average seen in previous work\cite{HaoWu2021}. This single scan was digitally filtered and serves to indicate the important characteristics of the Pc:PTP cooling signal; (i) its depth which was reported to be 50$^{+18}_{-32}$ K, and (ii) its duration of less than 0.5 ms.

\section{UV/Vis Absorption of NV$^-$ diamond sample}

\begin{figure*}[h!]
\includegraphics[width=0.5\textwidth]{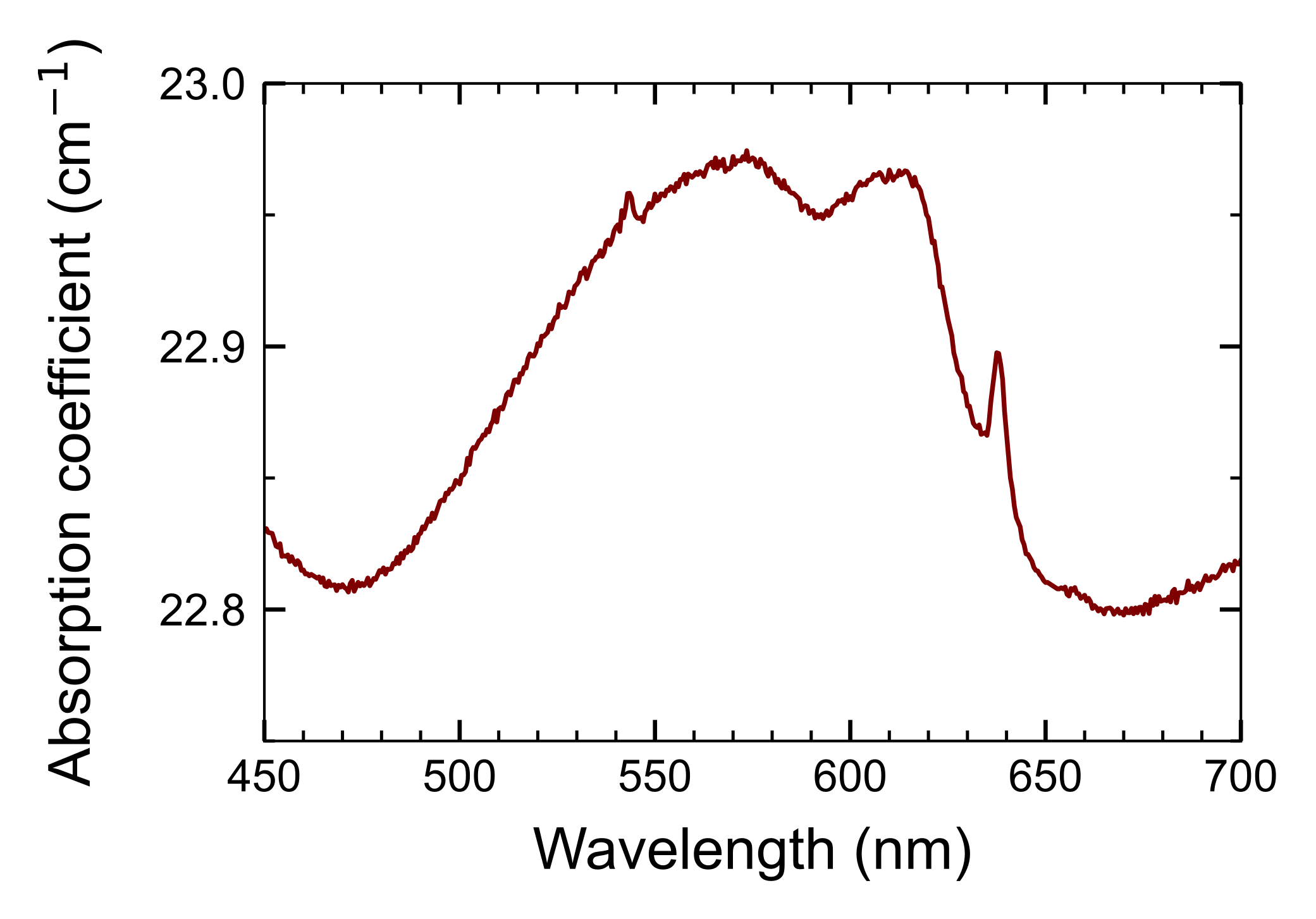}
\caption{\label{fig:abscoeff}Absorption coefficient measured through UV/Vis spectroscopy of the NV$^-$ diamond.}
\end{figure*}

The NV$^-$ diamond was placed in the beam path within an Agilent-Cary 5000 UV/Vis/NIR spectrophotometer, with its flat `table' (to use the nomenclature describing brilliant-cut diamonds) facing the light source. The zero-phonon line (ZPL) of NV$^-$ diamond is clearly visible as a sharp peak at 638 nm. The path length ($L$), which is governed by the total depth of the diamond, was 0.15 cm. After taking an initial measurement with the diamond placed in the light path, a second background measurement was taken with the diamond absent. After subtracting the background spectrum from the sample spectrum, the absorption coefficient could be calculated. We define absorbance $A$, absorption coefficient $\alpha$ and the Beer-Lambert equation as below:
\begin{eqnarray}
A=\log_{10}\frac{I_0}{I},\\
\frac{I}{I_0}=(1-R)^2\exp(-\alpha L) \label{beer},\\
R=\left|\frac{n_1-n_2}{n_1+n_2}\right|^2 \label{fresnel}.
\end{eqnarray}
The coefficient $(1-R)^2$ in the Beer-Lambert equation (Eq.~\ref{beer}) accounts for the reflection of light (at normal incidence) upon entering and exiting the sample. Since the diamond is a brilliant cut however, this is an approximation since the light may not be at normal incidence when entering or exiting the sample. Eq.~\ref{fresnel} is Fresnel's equation, where $n_1$ is the refractive index of air, and $n_2$ the refractive index of diamond (2.42). $A$ is measured directly as a function of wavelength from the spectrometer, and so through it $\alpha$ can be calculated and plotted as in Fig. S~\ref{fig:abscoeff}.
The concentration of NV$^-$ centers can be estimated using the absorption cross-section ($\sigma_\lambda$) of NV$^-$ centers at a specific wavelength. The number of NV$^-$ centers per unit volume of the sample would be given by $\alpha/\sigma_\lambda$. Using the absorption cross-section for the ZPL near 638 nm from literature ($4\times10^{-17}$~ cm$^{2}$)\cite{Fraczek17}, we estimate the concentration of NV$^-$ centers in our sample to be $6\times10^{17}$~cm$^{-3}$.
\section{NV$^-$ Diamond Spin Dynamics}
The spin dynamics of NV$^-$ diamond under optical pumping were simulated to produce the dotted plots for $T_{\text{mode}}$ seen in Fig.~3(b) and 4(b) in the main text. Referring to the Jablonski diagram from Fig.~1(a) in the main text, we define the coupled differential equations that govern the populations ($N$) of the energy levels ($\ket{0},\ket{1},\ket{2},\ket{3},\ket{4},\ket{5},\ket{S}$) and the number of thermal photons $q$.
\begin{eqnarray}
&\dot{N_{0}}=-(\xi+2\gamma_{02}+Bq)N_{0}+\gamma_{02}N_1+(\gamma_{02}+Bq)N_2+(\xi+k_{sp})N_3+k_{S0}N_S,\\
&\dot{N_{1}}=\gamma_{02}N_{0}-(\xi+\gamma_{02})N_{1}+(\xi+k_{sp})N_4+k_{S2}N_S,\\
&\dot{N_{2}}=(\gamma_{02}+Bq)N_{0}-(\xi+\gamma_{02}+Bq)N_{2}+(\xi+k_{sp})N_5+k_{S2}N_S,\\
&\dot{N_{3}}=\xi N_{0}-(\xi+k_{sp}+k_{3S})N_{3},\\
&\dot{N_{4}}=\xi N_{1}-(\xi+k_{sp}+k_{5S})N_{4},\\
&\dot{N_{5}}=\xi N_{2}-(\xi+k_{sp}+k_{5S})N_{5},\\
&\dot{N_{S}}=k_{3S}N_3+k_{5S}(N_4+N_5)-(k_{S0}+2k_{S2})N_S,\\
&\dot q=-\omega_{\text{mode}}\left(\frac{1}{Q_0}+\frac{1}{Q_{ex}}\right)\left(q-\frac{k_B}{hf_{\text{mode}}} T_0\right)+Bq(N_2-N_0).
\end{eqnarray}
The subscript of each population $N$ denotes which energy level it belongs to (e.g. $N_3$ is the population of the $\ket{3}$ level). $h$ and $k_B$ are Planck's constant and Boltzmann's constant respectively. The construction of the spin dynamics model will be discussed first, followed by detailed explanations for the two quantities $\xi$ (pumping parameter governing light excitation) and $Bq=W_{02}$ (stimulated transition rate governing microwave excitation). Under critical coupling, $Q_0=Q_{ex}$. The constants within the dynamics equations are summarized in Table~S\ref{tab:rateconstants} (ISC refers to intersystem crossing).
\begin{table}
\caption{\label{tab:rateconstants}Table of constants within the differential equations}
\begin{ruledtabular}
\begin{tabular}{lccr}
Description&Symbol&Value&Ref.\\
\hline
Spin-lattice relaxation rate (between $\ket{0}$ and $\ket{2}$) & $\gamma_{02}$ & $(1/0.012)\ s^{-1}$&based on experiment\\
Spontaneous emission rate\footnote{Averaged over all NV$^-$ orientations in Ref.~\onlinecite{Tetienne2012}} & $k_{sp}$ & $6.6\times10^7\ s^{-1}$&[\onlinecite{Tetienne2012}]\\
ISC rate ($\ket{S}\rightarrow\ket{0}$)$^{\text{a}}$ & $k_{S0}$ & $1.0\times10^{6}\ s^{-1}$&[\onlinecite{Tetienne2012}]\\
ISC rate ($\ket{S}\rightarrow\ket{2}$)$^{\text{a}}$ & $k_{S2}$ & $7.3\times10^{5}\ s^{-1}$&[\onlinecite{Tetienne2012}]\\
ISC rate ($\ket{3}\rightarrow\ket{S}$)$^{\text{a}}$ & $k_{3S}$ & $7.9\times10^{6}\ s^{-1}$&[\onlinecite{Tetienne2012}]\\
ISC rate ($\ket{5}\rightarrow\ket{S}$)$^{\text{a}}$ & $k_{5S}$ & $5.3\times10^{7}\ s^{-1}$&[\onlinecite{Tetienne2012}]\\
Mode frequency & $f_{\text{mode}}$ & $2.872\times10^9\ s^{-1}$&\\
Mode angular frequency & $\omega_{\text{mode}}$ & $2\pi(2.872\times10^9)\ s^{-1}$&\\
Mode unloaded quality factor & $Q_0$ & $2900\times2$&\\
Mode external quality factor & $Q_{ex}$ & $2900\times2$&\\
Ambient temperature of cavity (room temp.) & $T_0$ & $290$ K&\\
\end{tabular}
\end{ruledtabular}
\end{table}

From the Jablonski diagram in Fig.~1(a) in the main text, it is assumed that the following spin-lattice relaxation times are identical, and that the upward and downward spin-lattice relaxation times are also identical (this latter assumption is valid for experiments at high ambient temperature\cite{Hintze2017}): $\gamma_{0\rightarrow2}=\gamma_{0\rightarrow1}=\gamma_{1\rightarrow0}=\gamma_{2\rightarrow0}$. Here, arrows have been added to the subscripts to make the direction of transition more clear. Furthermore, it is assumed that there is no spin-lattice relaxation between the nearly-degenerate triplet sublevels $\ket{1}$ and $\ket{2}$: $\gamma_{12}=0$. Hence, only one parameter for the spin-lattice relaxation $\gamma_{02}$ is needed. $\gamma_{02}$ was varied during simulations to find a suitable value that emulates the decay time of the cooling signals from experiments. The ISC rates are assumed to be identical for nearly-degenerate triplet sublevels (i.e. $k_{S2}=k_{S1}$ and $k_{5S}=k_{4S}$).

The derivation for the differential equation of $q$ is the same as in previous work\cite{HaoWu2021}, where we again assume that the temperature of the internal `spin' system $T_{02}=hf_{02}/[2k_B\tanh^{-1}((N_0-N_2)/(N_0+N_2))]$ is negligible due to the high spin polarization of NV$^-$, and so the term is removed from the equation.

$\xi$ is the pumping parameter as used in previous work\cite{HaoWu2021}, which is defined as:
\begin{equation}
    \xi(t)=\frac{\lambda_P\sigma_{\lambda_P}}{hcA_Pl\alpha}(1-\exp(-l\alpha))(1-R)P(t)
\end{equation}
$\lambda_P=532$ nm is the laser pump wavelength, $\sigma_{\lambda_P}$ is the absorption cross-section of the NV$^-$ centers at the wavelength $\lambda_P$ ($3.1\times10^{-21}$ m$^2$)\cite{Tse-Luen2007}. $c$ is the speed of light. $A_P$ is the cross-sectional area of the pump beam incident on the sample ($1.76\times10^{-6}$ m$^2$), where the beam spot was a 1.5 mm diameter circle. $l$ is the thickness of the crystal as traversed by the pump beam (0.0015 m, assuming it is the depth of the diamond). $\alpha$ is the absorption coefficient at $\lambda_P$, which from Fig. S~\ref{fig:abscoeff} is $2.3\times10^3$ m$^{-1}$. A factor of $1-R$ is included to account for Fresnel reflection (as defined in Eq.~\ref{fresnel}) at the diamond surface, again assuming normal incidence (which is a simplification). $P(t)$ is the instantaneous pump power of the laser (in watts) as a function of time. For the simulation, a square pulse with 2 W peak power and 2 ms pulse width was used for simulating $T_{\text{mode}}$ in Fig. 3(b) of the main text, while a square pulse with 2 W peak power and 10 ms pulse width was used for simulating $T_{\text{mode}}$ in Fig. 4(b) of the main text. 

$Bq$ is the stimulated transition rate $W_{02}$ between $\ket{0}$ and $\ket{2}$, which is defined similarly in previous work\cite{HaoWu2021}. The equation is:
\begin{equation}
    W_{02}=Bq=\left(\frac{\mu_0\gamma^2hf_{\text{mode}}T_2^*\braket{\sigma^2}\eta_{\text{fill}}}{2V_{\text{mode}}}\right)q
\end{equation}
where $B$ is Einstein's $B$ coefficient (evaluated in Ref.~\onlinecite{Salvadori2017} based on the derivation in Ref.~\onlinecite{siegman1964microwave}) for the $\ket{0}\rightarrow\ket{2}$ transition. $\mu_0$ is the permeability of free-space, $\gamma\sim2\pi\times28$ GHz/T is the reciprocal gyromagnetic ratio for the $\ket{0}\rightarrow\ket{2}$ transition, $f_{\text{mode}}=2872$ MHz is the mode frequency, $T_2^*$ is the `inhomogeneous' spin-spin relaxation time, $\braket{\sigma^2}=0.5$ is the normalized transition probability matrix element for an allowed transition between two triplet sublevels (Table 5-1 in Ref.~\onlinecite{siegman1964microwave}),
$\eta_{\text{fill}}$ is the filling factor of the sample in the mode, and $V_{\text{mode}}$ is the mode volume. 

$T_2^*$ can be estimated upon assuming that dipolar coupling between substitutional nitrogen defects and NV$^-$ centers is the dominant mechanism for dephasing (lowering $T_2^*$)\cite{Barry2020}. The concentration of substitutional nitrogen defects in the sample often persists at the same concentration as (or higher than) the concentration of NV$^-$ centers\cite{Barry2020}, and so an NV$^-$ concentration of $6\times10^{17}$ cm$^{-3}$ would give $T^*_2\sim3\ \mu$s approximately\cite{Barry2020}. $\eta_{\text{fill}}$ and $V_{\text{mode}}$ have the equations below,
\begin{eqnarray}
\eta_{\text{fill}}=\frac{\int_{\text{ex. sample}}|\textbf{H}|^2\ dV}{\int_{\text{mode}}|\textbf{H}|^2\ dV}
\label{eq:fill},\\
V_{\text{mode}}=\frac{\int_{\text{mode}}|\textbf{H}|^2\ dV}{\text{max}[|\textbf{H}|^2]}
\end{eqnarray}
\textbf{H} is the magnetic field strength resulting from the TE$_{01\delta}$ mode, while $\int_{\text{ex. sample}}...\ dV$ and $\int_{\text{mode}}...\ dV$ indicate integration over and around the volume of the sample excited by the light and the entire mode bright spot respectively. $\text{max}[...]$ denotes the maximum value (scalar) of the functional argument. The filling factor reflects that only the volume of the NV$^-$ diamond excited by the light will be available to interact with the microwave mode and perform cooling. In the previous work, this factor was assumed to be unity and hence not included\cite{HaoWu2021}. Integrals for $\eta_{\text{fill}}$ and $V_{\text{mode}}$ can easily be calculated using the finite element method in COMSOL, and so the two quantities were estimated to be $\eta_{\text{fill}}\approx0.018$ and $V_{\text{mode}}\approx0.084$ cm$^3$. 

The change of $T_2$ to $T_2^*$ and the factor of $\eta_{\text{fill}}$ are two changes to the $B$ coefficient seen in Ref.~\onlinecite{HaoWu2021}. The use of $T_2^*$ is due to the presence of many inhomogeneous broadening mechanisms when measuring bulk NV$^-$ diamond\cite{Barry2020}. The stimulated transition rates would then be governed more by the inhomogeneous linewidth ($T_2^*$) rather than the homogeneous linewidth (refer to Chap~5, page~222 of Ref.~\onlinecite{siegman1964microwave} on homogeneous vs. inhomogeneous broadening).

The initial conditions used in order to solve the differential equations are as follows. Within the volume of the NV$^-$ diamond that is excited by the light, there exists a certain amount of spins. Firstly, this volume is a cylinder of diameter 1.5 mm and height of 1.5 mm (as defined by $A_P$ and the crystal thickness $l$ traversed by the pump beam), which is admittedly a simplification considering the brilliant-cut of the diamond. Using the estimated concentration of NV$^-$ centers ($6\times10^{17}$ cm$^{-3}$), we initially calculated the number of spins in that volume participating in the dynamics as $N_T=1.6\times10^{15}$, where $N_T$ denotes the total number of (participating) spins. However, initial simulations using this value gave a $T_\text{mode}$ that was much deeper than that experimentally observed using the power reduction measurements, possibly caused by $N_T$ being overestimated. $N_T$ was then scaled down to the value of $N_T=0.72\times10^{15}$, which more accurately reproduced the cooling depth. The discrepancy could be due to the approximate method of measuring the UV-Vis spectrum, which would alter the concentration and absorption coefficient values. The initial populations $N_0,N_1,N_2$ before pumping would then be based on the Boltzmann distribution at $T_0=290$~K. This gives approximately $N_0=0.3334N_T,N_1=0.3333N_T,N_2=0.3333N_T$. All other initial populations are assumed to be negligible ($N_3=N_4=N_5=N_S=0$), and the initial amount of thermal photons is calculated using $q=\left(\exp(hf_{\text{mode}}/k_BT_{0})-1\right)^{-1}$, which gives $q=2103$.

\begin{figure*}[ht!]
\includegraphics{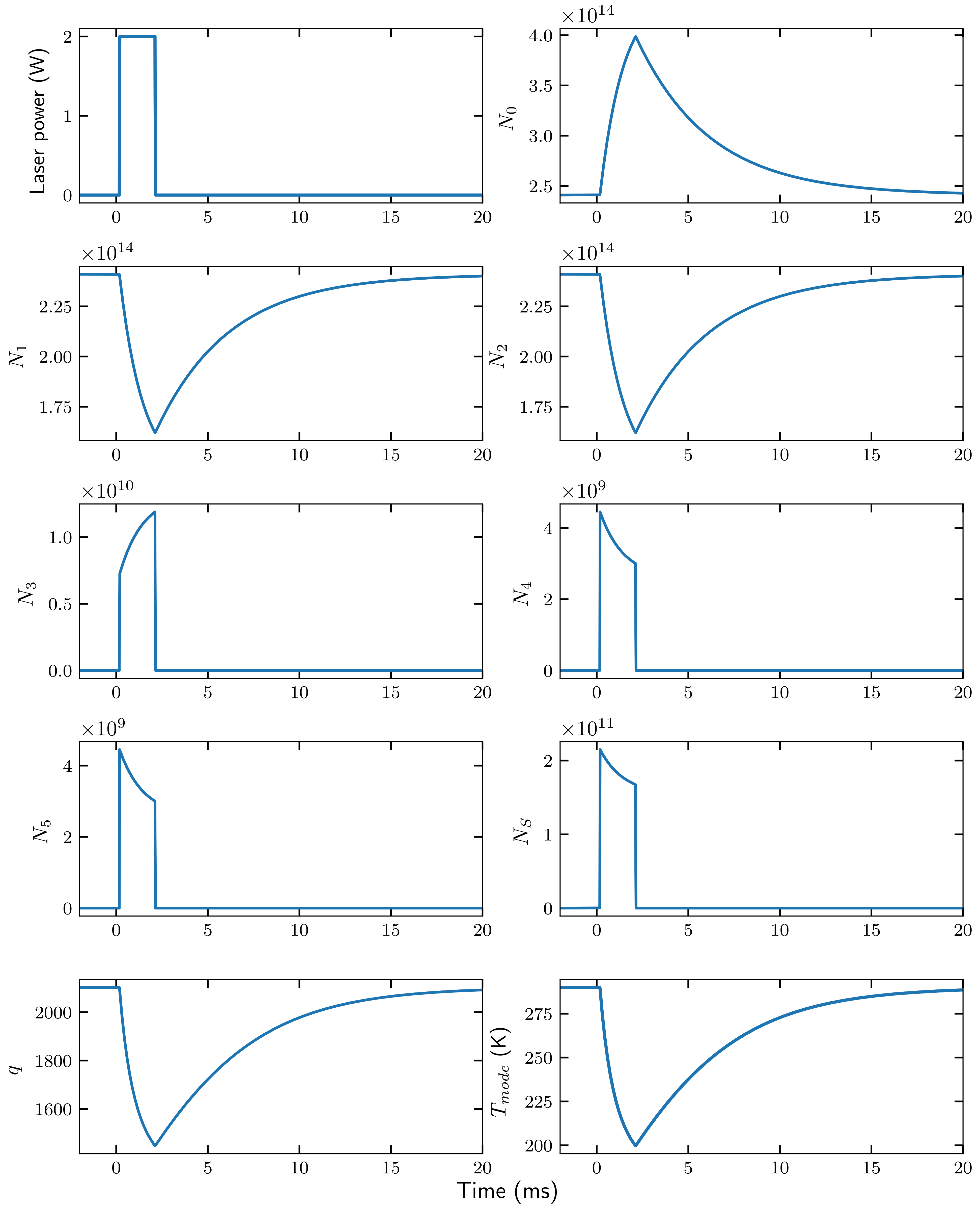}
\caption{\label{fig:2ms}Spin dynamics simulation using 2 ms laser pulse for $P(t)$.}
\end{figure*}

\begin{figure*}[ht!]
\includegraphics{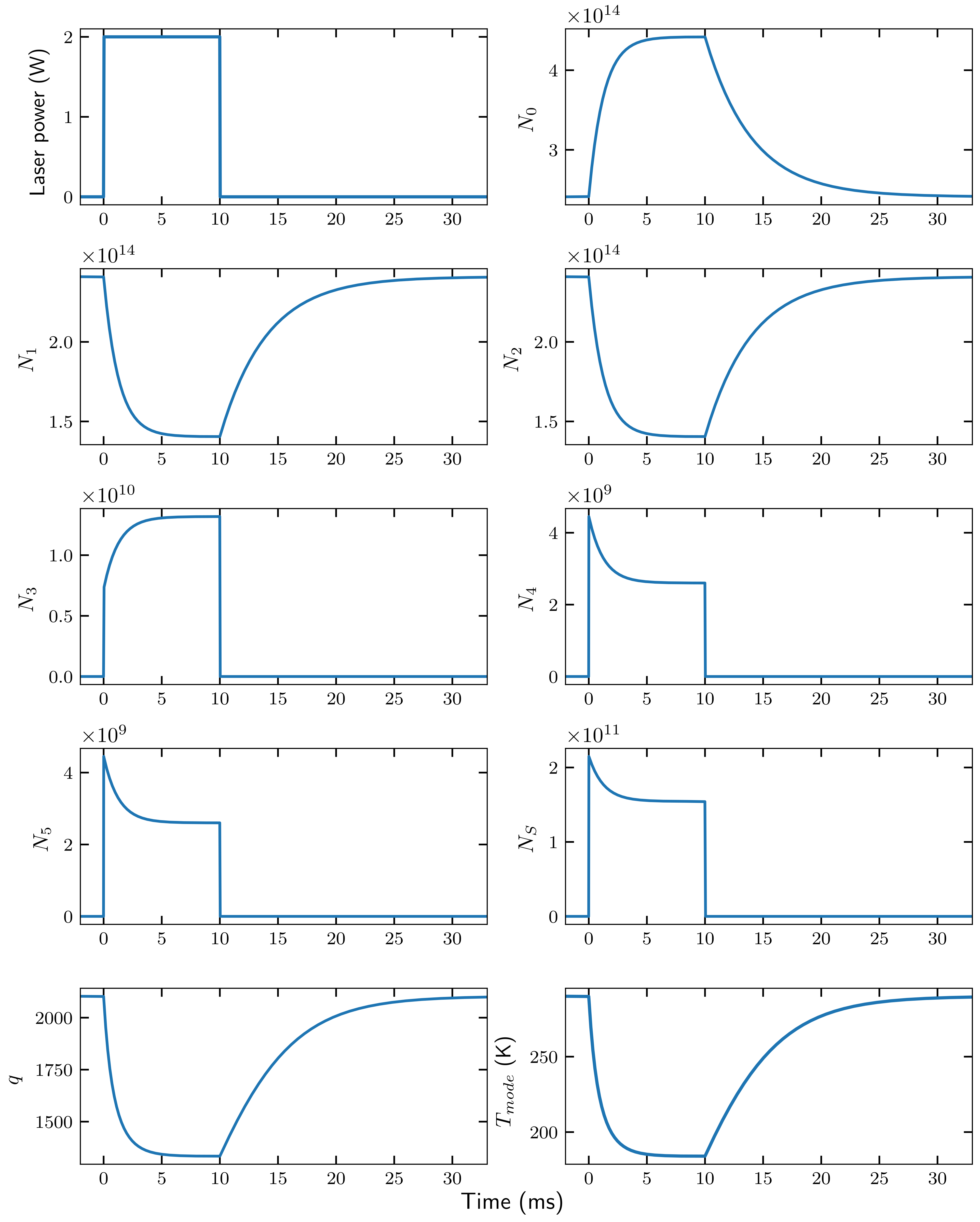}
\caption{\label{fig:10ms}Spin dynamics simulation using 10 ms laser pulse for $P(t)$.}
\end{figure*}

With the initial parameters calculated, the differential equations were solved using Python (specifically, the \textit{scipy.integrate.odeint} function which imports the LSODA method from FORTRAN). The laser pulse was defined through $P(t)$ in $\xi(t)$, which was included as a time-dependent parameter in the differential equations. The system was first allowed to evolve and equilibrate after a short period of time ($\sim 1$ ms) with $\xi(t)=0$, since the coupling of the spins to the mode (through $W_{02}$) will cause a very small perturbation of the initial populations even under no illumination. Then, the system was evolved with the laser pulse (via $\xi(t)$) turned on. Plots for the evolution of the populations (short equilibration period not shown) are given in Fig.~S~\ref{fig:2ms} (when using a 2-ms pulse) and S~\ref{fig:10ms} (when using a 10-ms pulse).

The resulting simulated $q$ can be used to simulate $T_\text{mode}$ through inverting Eq. 1 in the main text. The $T_{\text{mode}}$ plots in Fig.~S~\ref{fig:2ms} and S~\ref{fig:10ms} are then the simulated $T_{\text{mode}}$ plots used in Fig.~3(b) and 4(b) in the main text.

\newpage
\newpage
\newpage

\section{Noise analysis}
In-depth explanation of the noise theory has already been done in previous work\cite{HaoWu2021}, and so we will concentrate on the primary equation used to relate the noise power reduction $\Delta P$ to $T_{\text{mode}}$\cite{HaoWu2021}:
\begin{equation}
    \Delta P(T_{\text{mode}})=10\log_{10}\frac{G_{\text{LNA}}\left[(T_{\text{min}}+T_{\text{mode}})(1-|\Gamma_c|^2)+4T_0\frac{R_n|\Gamma_c-\Gamma_{\text{opt}}|^2}{Z_0|1+\Gamma_{\text{opt}}|^2}+T_{\text{image}}\right]+T_{\text{REC}}}{G_{\text{LNA}}\left[(T_{\text{min}}+T^0_{\text{mode}})(1-|\Gamma^0_c|^2)+4T_0\frac{R_n|\Gamma^0_c-\Gamma_{\text{opt}}|^2}{Z_0|1+\Gamma_{\text{opt}}|^2}+T_{\text{image}}\right]+T_{\text{REC}}}\label{deltap}
\end{equation}
\begin{equation}
    \Gamma_c=\frac{T_{\text{mode}}}{T_0}-1
\end{equation}
The normalization impedance is $Z_0=50\ \Omega$, initial mode noise temperature $T^0_{\text{mode}}$ is the same as $T_0=290$ K, and the initial cavity reflection coefficient $\Gamma_c^0=0$ as the cavity was initially critically coupled. It is again assumed that the connection between the cavity and the first LNA, which was simply the length of the microwave iris together with a cable ($\sim8$ cm length in total), was sufficiently short so that its phase length and thermal (Johnson) noise could be ignored.
The first LNA was a Qorvo QPL9547EVB-01. The noise parameters appearing in Eq.~\ref{deltap} were taken from this LNA's datasheet (interpolating to the 2872 MHz operating frequency used where possible). The remaining constants were either measured (through using the VNA) or calculated. Table~\ref{tab:powerconstants} displays the values used. Further details on the origin and meaning of each parameter are provided in previous work\cite{HaoWu2021}.
\begin{table}
\caption{\label{tab:powerconstants}Table of constants for the noise reduction equation}
\begin{ruledtabular}
\begin{tabular}{lccr}
Description&Symbol&Value&Ref.\\
\hline
Power gain of first LNA (linear units) & $G_{\text{LNA}}$ & 32.5&measured\\
LNA minimum noise temperature & $T_{\text{min}}$ & 17.4 K&[\onlinecite{qorvoamp}]\\
LNA noise resistance & $R_n$ & 1.1 $\Omega$&[\onlinecite{qorvoamp}]\\
Optimum source reflection coefficient & $\Gamma_{\text{opt}}$ & $-0.131+0.189i$&[\onlinecite{qorvoamp}]\\
Image noise temperature & $T_{\text{image}}$ & 25.5 K&calculated using equation in Ref.~\onlinecite{HaoWu2021}\\
Noise temperature of the rest of the receiver & $T_{\text{REC}}$ & 43 K& calculated using Friis formula\cite{kraus1986}
\end{tabular}
\end{ruledtabular}
\end{table}

Eq.~\ref{deltap} is used to generate the curve in Fig.~3(c) in the main text. The equation was then numerically inverted to produce the inverse function that gives values of $T_{\text{mode}}$ when given a certain value of $\Delta P$. The experimentally measured $\Delta P$ can then be fed into the inverse function to give experimental values of $T_{\text{mode}}$ seen in Fig.~3(b) and 4(b).